\input psfig
\documentstyle[11pt]{article}

\begin{document}
\vspace*{-1.25in}
\small{
\begin{flushright}
FERMILAB-PUB-96/148-T \\[-.1in] 
hep-lat/9607032n \\[-.1in] 
June, 1996 \\
\end{flushright}}
\vspace*{.85in}
\begin{center}
{\Large{\bf Electromagnetic Structure of Light Baryons \\
 in Lattice QCD \\}
}
\vspace*{.45in}
{\large{A.~Duncan$^1$\footnote{Permanent address: 
Dept. of Physics and Astronomy, 
University of Pittsburgh, Pittsburgh, PA 15620.}
, E.~Eichten$^2$ and H.~Thacker$^3$}} \\ 
\vspace*{.15in}
$^1$Dept. of Physics, Columbia University, New York, NY 10027 \\
$^2$Fermilab, P.O. Box 500, Batavia, IL 60510 \\
$^3$Dept.of Physics, University of Virginia, Charlottesville, VA 22901
\end{center}
\vspace*{.3in}
\begin{abstract}
A method for computing electromagnetic properties of hadrons 
in lattice QCD is applied to the extraction of electromagnetic properties
of the octet baryons.
This allows a determination of 
the full dependence of the baryon masses on 
the charges and masses of the valence quarks. 
Results of a first numerical study (at $\beta=5.7$ with Wilson action 
and light quark masses fixed from the pseudoscalar meson spectrum)
are reported.  
The octet baryon isomultiplet splittings 
(with statistical errors)
are found to be:
${\rm N} - {\rm P} = 1.55(\pm 0.56)$, 
$\Sigma^0 - \Sigma^+ = 2.47(\pm 0.39)$, 
$\Sigma^- - \Sigma^0 = 4.63(\pm 0.36)$ and  
$\Xi^- - \Xi^0 = 5.68(\pm 0.24)$ MeV. 
Estimates of the systematic corrections arising from finite volume 
and the quenched approximation are included in these results.
\end{abstract}


\newpage


The accurate determination of the neutron-proton mass difference
represents one of the thorniest and longest standing problems of hadronic physics.
An early review of Zee \cite{Zee} concludes with 19 appendices outlining various
approaches to the problem. The advent of modern quantum chromodynamics (QCD)
has dissipated  a great deal of the conceptual fog  
surrounding this problem, but we are still faced with the difficult
technical problem of computing accurately the competing effects of the up-down
quark mass difference (the ``tadpole", in ancient jargon) and the photon cloud
energy in a complicated hadronic state.
Since the contribution to hadronic
mass splittings within isomultiplets from  
virtual photon effects 
is comparable to the size of the up-down quark  mass difference, 
an accurate computation of isospin splittings
requires the inclusion of electromagnetic effects in the context of
nonperturbative QCD dynamics. In this letter, we apply a method recently
used \cite{DET} to extract the electromagnetic contributions to pseudoscalar masses
to the problem of the octet baryon spectrum.
In addition to the SU(3) color field, a U(1) electromagnetic field on the
lattice is introduced and treated by Monte Carlo methods. The resulting SU(3)$\times$U(1)
gauge configurations are then analyzed by standard hadron propagator techniques. 
One of the main results of our earlier work
\cite{DET} was to demonstrate that calculations
done at larger values of the quark electric charges (roughly 2 to 6 times
physical values)
lead to accurately measurable hadronic isospin splittings,
while still allowing  perturbative extrapolation to  physical values. A 
useful result of that work was the extraction of up, down and strange quark
bare lattice masses appropriate for a lattice of quenched configurations
at $\beta$=5.7 (and Wilson action), which can serve as input  
to a computation of the baryonic isomultiplets. 

The strategy of the present calculation is as follows.  Quark 
propagators are generated in the
presence of Coulomb gauge background SU(3)$\times$U(1) fields described above.
(A detailed description of our formulation of the U(1) field is presented 
in our previous work\cite{DET}.)
Quark propagators are calculated for a variety of electric charges
and light quark mass values, and with either a local or smeared source.
187 gauge configurations, separated by 1000 Monte Carlo sweeps,
were generated at $\beta=5.7$ on a $12^3\times 24$ lattice.
We have used four different values of charge 
given by $e_q =$0, -0.4, +0.8, and -1.2 in units in which the electron charge
is $e=\sqrt{4\pi/137} =.3028\ldots\;$. 
For each quark charge we calculate propagators for
three light quark mass values in order to allow a chiral extrapolation. 
From the resulting 12 quark propagators, 936 different octet baryon
three-quark combinations can be formed. 
The baryon propagators are then computed and masses for all 936 
states extracted. 

 The analysis then proceeds by a combination
of chiral and QED perturbation theory. 
In quenched QCD it is known \cite{sharpe} that 
baryon masses are described by a function
of the bare quark masses involving not only linear but also
nonanalytic $m_{q}^{3/2}$ terms, as well as terms involving logarithms
of the quark mass arising from the same hairpin diagrams
familiar in the quenched meson spectrum \cite{bern,sharpmes}.
The latter terms arise from the hairpin diagrams associated with 
unsuppressed $\eta^{\prime}$ loops in the quenched approximation.
They appear to be extremely small numerically for the light quark masses
we consider\cite{hank}, and are neglected in our baryon analysis. However,
we do include the terms of order $m_q^{3/2}$.
Thus a general octet baryon mass is written in an expansion of the form

\begin{equation}
\label{eq:ChPT}
m_{B} = A(e_{q1}, e_{q2}, e_{q3}) + \sum_{i} m_{qi}B_i(e_{q1}, e_{q2}, e_{q3})
 \\ 
+ \sum_{i,j}(m_{qi}+m_{qj})^{3/2}C_{ij}(e_{q1}, e_{q2}, e_{q3}) 
\end{equation}
where $e_{q1}, e_{q2}, e_{q3}$ are the three quark charges, and 
$m_{q1}, m_{q2}, m_{q3}$
are the three bare quark masses, defined in terms of the Wilson hopping parameter by
$(\kappa^{-1} - \kappa^{-1}_c)/2a$. (Here $a$ is the lattice spacing.)
Each of the coefficients $A,B_i,C_{ij}$ in (\ref{eq:ChPT}) is then expanded
in powers of the quark charges $e_{q1}, e_{q2}, e_{q3}$, with terms up to fourth
 order for $A$, second order for $B_i$, and with no charge dependence 
 assumed for the nonanalytic $C_{ij}$ terms.
Because of the electromagnetic self-energy shift, 
the value of the critical hopping parameter must be determined
independently for each quark charge. This is done \cite{DET} 
by requiring that the mass of
the neutral pseudoscalar meson vanish at $\kappa=\kappa_c$. The results obtained
in \cite{DET} for the $\kappa_c$ values for various electric charge are inputs
to our baryon calculations (as are the light quark masses extracted from the
pseudoscalar meson spectrum). The fitting formula (\ref{eq:ChPT}) turns out to 
have 30 parameters once all symmetries are exploited.

  The success of the procedure outlined above clearly depends on the accurate
extraction of the baryon masses for a large class of baryon states built from
quarks of varying mass and electric charge. As in the meson studies
of our previous work\cite{DET}, quark
propagators were calculated at hopping parameters ($\kappa$) 0.161, 
0.165 and 0.1667 
(with $\beta$ = 5.7) for charge zero quarks. For the nonzero charge quarks,
the hopping parameters were shifted by an amount computed from improved
lattice perturbation theory in order to keep the quark masses for
nonzero electric charge close to their values at zero charge\cite{DET}. 
A preliminary
study using both single and multistate fits and employing local and smeared
sources and sinks indicated that stable effective mass plots with reasonably
small statistical errors could only be obtained from correlators with 
smeared sources and local sinks. A single quark propagator
smearing function (obtained from
a lattice relativistic quark model with parameters chosen to fit a pion
wavefunction) was used throughout, for both the meson and baryon studies,
so it is not surprising to find that the systematic errors induced by
higher states were minimal in a limited range of baryon masses. In 
particular, baryons with masses (in lattice units) between 1.10 and 1.30
were found to give sizable effective mass plateaux with fairly small
statistical errors (see Fig.1). The errors are obtained from single-elimination
jackknife averages using 187 gauge configurations, each 
separated by 1000 Monte Carlo
sweeps.
    
\begin{figure}
\psfig{figure=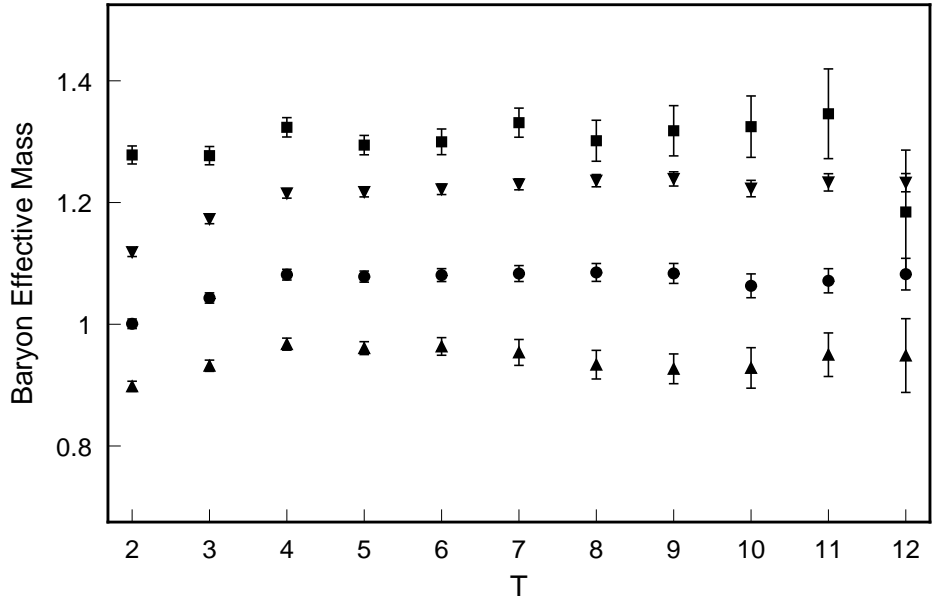,
width=0.95\hsize}
\caption{Some typical effective mass plots from smeared-local baryon 
correlators. Effective masses and times in lattice units.}
\label{fig:effmass}
\end{figure}

  Octet baryon states correspond to a mixed spin-flavor symmetry, so
 correlators were computed for two distinct spin-flavor combinations.
  In the first combination, the first two quarks are coupled to 
 spin and isospin 1 - as sigma hyperons can only be constructed in this
 way, this combination is called ``sigma-type". In the second combination
 (dubbed ``lambda-type"),
 the first two quarks are coupled to spin and isospin 0. Isospin-$\frac{1}{2}$
 states (nucleon and $\Xi$) can be constructed from either the sigma or
 lambda type spin-flavor combinations. Although the isospin-$\frac{1}{2}$
 masses determined from either combination must agree in the full theory
 in the continuum limit, quenched and finite lattice spacing corrections
 will lead to discrepancies in simulations on small lattices. Since we 
 are only trying to determine the splittings within a given isomultiplet,
 we have
 chosen to determine the scale in all cases by fixing the average nucleon mass 
 at its physical value.  For the sigma-type combinations, our $\beta$=5.7 
 configurations (with unimproved action) then lead to a scale of about 
 $a^{-1}$=1.37 GeV, while the scale for lambda-type combinations is
 found to be about $a^{-1}$=1.33 GeV. 

   As mentioned previously, we find the cleanest effective-mass plateaux
 in a limited range of baryon masses, presumably because of the fixed 
 smearing wavefunction used for all the quark propagators. It should be
 emphasized that even a limited mass range for the baryons still contains
 baryons with widely ranging quark masses and charges, so the fitting
 formula (\ref{eq:ChPT}) can be used to extract the $A, B_i, C_{ij}$
 coefficients, allowing an extrapolation to physical values of quark mass
 and electric charge. We have further restricted the portion of the
 baryon spectrum being fitted by varying the baryon mass window (for each
 choice of Euclidean time window used to extract a mass) until the 
 $\chi^2/{\rm dof}$ was minimal. For example, using a Euclidean time
 window from $t=$5 to $t=$8, the mass window (lattice units) from 
 1.20 to 1.26 was found to contain 74 sigma-type baryons. Determining the 30
 parameters in (\ref{eq:ChPT}) by fitting this set of masses gave
 a $\chi^2/{\rm dof}$ of 1.33 . By contrast, using the mass window
 from 1.15 to 1.20 (122 baryons), the chi-square fit minimizes at
 $\chi^2/{\rm dof}$=2.16. For each choice of Euclidean time window, we
 have performed the fit to (\ref{eq:ChPT}) using a baryon mass window
 which optimizes the $\chi^2/{\rm dof}$. For the lambda-type baryons,
 the best fit was obtained for the time window $t=$5 to $t=$9 using 54
 baryons in the mass range 1.21 to 1.26, giving a $\chi^2/{\rm dof}$ of
 1.40.

   The chi-squared fit to the chiral and electromagnetic expansion
 formula gives a central value for each fitting parameter, and a
 full 30x30 covariance matrix specifying the correlated errors on
 the fit parameters. With this information, one can determine the
 mass of any given octet baryon by extrapolating to physical values
 of quark mass and charge, together with an error which properly 
 includes the statistical correlations among the fit parameters.
 The propagators for different electric charge are highly correlated,
 so it is not surprising to find that the statistical error on the
 center of gravity of baryon isomultiplets is considerably larger
 than the error on splittings within multiplets (typically, two
 orders of magnitude). This is apparent in Table 1, where the raw
 lattice results for the baryon octet spectrum with statistical errors,
  but no corrections for finite volume or quark-loop
 effects,  are given for three
 different choices of Euclidean time fitting window. Corrections
 due to these systematic effects will be estimated and included below. 

\begin{table}
\begin{center}
\caption{Raw lattice results for baryon octet 
($\beta$=5.7, 12$^3$x24, 187 configurations). Quark masses are
the bare lattice masses for Wilson action. All results are in MeV.}
\begin{tabular}{|c|c|c|c|}
\hline
\multicolumn{1}{|c|}{Baryon State}
&\multicolumn{1}{c|}{Window 5-8}
&\multicolumn{1}{c|}{Window 6-9}
&\multicolumn{1}{c|}{Window 5-9}   \\ 
\multicolumn{1}{|c|}{~~}
&\multicolumn{1}{c|}{($\chi^2/{\rm dof}$=1.33)}
&\multicolumn{1}{c|}{($\chi^2/{\rm dof}$=1.51)}
&\multicolumn{1}{c|}{($\chi^2/{\rm dof}$=1.57)}   \\ \hline
\multicolumn{1}{|c|}{Parameters}
&\multicolumn{1}{c|}{$a^{-1}=1370$}
&\multicolumn{1}{c|}{$a^{-1}=1370$}
&\multicolumn{1}{c|}{$a^{-1}=1280$} \\ 
\multicolumn{1}{|c|}{$m_{u,d,s}$} 
&\multicolumn{1}{c|}{(3.57,7.10,155)}
&\multicolumn{1}{c|}{(3.57,7.10,155)}
&\multicolumn{1}{c|}{(3.88,7.54,180)} \\ \hline
N  &  935.92$\pm$42.4  & 952.24 $\pm$ 43.6 &  942.13 $\pm$ 36.7  \\
P  &  933.07 $\pm$42.9   & 949.42 $\pm$ 44.0     &  940.00 $\pm$ 37.1 \\  
N-P & 2.83 $\pm$ 0.56  & 2.82 $\pm$ 0.57 &  2.13 $\pm$ 0.50 \\ 
$\Sigma^{+}$  & 1171.6 $\pm$ 25.6  & 1181.4 $\pm$ 26.1 &  1181.5 $\pm$ 19.2 \\
$\Sigma^{0}$  & 1175.1 $\pm$ 25.3  & 1185.0 $\pm$ 25.8 &  1184.4 $\pm$ 18.9 \\
$\Sigma^{-}$  & 1179.1 $\pm$ 25.0  & 1189.1 $\pm$ 25.5 &  1187.8 $\pm$ 18.7 \\
$\Sigma^{0}-\Sigma^{+}$  & 3.43 $\pm$ 0.39  & 3.56 $\pm$ 0.38 &  2.92 $\pm$ 0.34 \\
$\Sigma^{-}-\Sigma^{0}$  & 4.04 $\pm$ 0.36  & 4.13 $\pm$ 0.43 &  3.36 $\pm$ 0.33 \\
$\Sigma^{+}+\Sigma^{-}-2\Sigma^{0}$  & 0.61 $\pm$ 0.19 & 0.57 $\pm$ 0.26 & 0.44 $\pm$ 0.19\\
$\Xi^{-}$  &  1312.9 $\pm$ 14.5 & 1317.7 $\pm$  15.9 &  1319.1 $\pm$ 10.9\\
$\Xi^{0}$  &  1308.2 $\pm$ 14.6 & 1312.7 $\pm$ 16.1 &  1314.8 $\pm$ 11.0\\
$\Xi^{-}-\Xi^{0}$  &  4.72 $\pm$ 0.24 & 5.01 $\pm$ 0.33 &  4.30 $\pm$ 0.24 \\
$\Lambda^{0}$  & 1098 $\pm$ 52  & 1107 $\pm$ 22 &  1104 $\pm$ 69\\
\hline
\end{tabular}
\end{center}
\end{table}

In the second row of Table 1 we indicate the quark mass parameters and lattice
scale assumed in generating the sigma-type masses for each of the fitting window choices.
The lattice scale has been fixed in each case by requiring the nucleon center
of gravity to sit at (roughly) the physical value. The masses
within isomultiplets are highly correlated, so that the errors on the splittings
are far smaller than on the multiplet center of gravity. This is expected
since all the masses are calculated on the same set of gauge configurations.
Once the lattice scale is
fixed, we input up and down quark masses as previously determined \cite{DET} by
analysis of the pseudoscalar meson spectrum (these will depend on the scale).
The strange quark mass is known to fall at a higher value when determined from
the baryon spectrum (a discrepancy which will presumably disappear on removal
of finite $a$ and quenched errors), so we have chosen to fix it using the
center of gravity of the $\Xi$ hyperon, which has the smallest statistical errors
in our analysis. The center of gravity of the $\Sigma$ multiplet and the $\Lambda$
mass are then predictions of the analysis. The variations between the columns of
Table 1 may be taken as an indication of the size of systematic errors associated
with our choice of fitting window, which are generally smaller than the  statistical 
errors.  (Our bare quark masses correspond to an unimproved Wilson action at
$\beta$=5.7, and are significantly larger than the continuum 
extrapolated $\bar{\rm MS}$ values.)

 For the remainder of this letter, we focus our attention on the splittings within
isospin multiplets. Although the statistical errors are much smaller here, the
relative impact of systematic errors 
(primarily those induced by finite volume, elimination 
of virtual quark loops, and lattice discretization) is far more important. To
simplify the discussion, we shall henceforth restrict ourselves to the choice
of Euclidean time-window 5-8 (which moreover gives the smallest $\chi^2$), i.e. the
second column of Table 1. By phenomenological estimates, discussed below, 
we find that the finite volume
errors amount in all cases to shifts of less than 1 MeV, and are therefore totally
irrelevant to the discussion of the isomultiplet centers of gravity. We have
estimated and included these finite volume corrections 
(as well as quenched corrections to
 splittings) below-
 the corrections due to lattice discretization and finite volume
will be studied directly in an upcoming run employing improved actions and 
larger lattices.

Given the presence of massless physical degrees of freedom we expect finite 
 volume effects which fall only as inverse powers of the lattice size. These
corrections can be studied directly on the lattice by repeating the calculations
on lattices of varying physical volume. For the present, we estimate them by
using the known dominance \cite{GassLeut} of the Born contribution to the
dispersive evaluation of the Cottingham \cite{Cott} formula. The self-energy
 shift of a hadronic state
 arising from single photon exchange can be written as a sum of an electric
and magnetic contribution, which on a spatial LxLxL lattice take the form
\begin{eqnarray}
\label{eq:finvol1}
 \delta m_{\rm el} &=& 2\pi\alpha m \frac{1}{L^3}\sum_{\vec{q}\neq 0}
 \frac{G_{E}(q)^2}{|q|}
 \{\frac{2}{q^2 +4m^2}+\frac{1}{2m^2}(\sqrt{1+\frac{4m^2}{q^2}}-1)\} \\
\label{eq:finvol2}
 \delta m_{\rm mag} &=& -\frac{\pi\alpha}{2m^3}\frac{1}{L^3}\sum_{\vec{q}\neq 0}
 |q|G_{M}(q)^2 \{\sqrt{1+\frac{4m^2}{q^2}}-1-\frac{1}{2}\frac{1}{1+\frac{q^2}
{4m^2}}\}
\end{eqnarray}
where the momentum vectors $\vec{q}$ are the appropriately discretized bosonic
momenta for the finite LxLxL lattice. The elimination of the zero mode
in the noncompact formulation \cite{DET} of the electromagnetic field we employ 
means that the $\vec{q}=0$ term is omitted in (\ref{eq:finvol1},\ref{eq:finvol2}).
The finite volume error can then be
extracted by studying the dependence of (\ref{eq:finvol1},\ref{eq:finvol2}) on L, holding the 
lattice spacing fixed. The charged baryon states contain a monopole 
contribution which falls as 1/L in the infinite volume limit. Using a dipole
Ansatz \cite{leutrev} for the electric and magnetic form factors, one
finds a positive shift $\delta m_{\rm el}$= 0.84 MeV for the proton in going
from L=12 to L=$\infty$. This electric shift is essentially the same for all
charged members of the octet (a static approximation is very good for this
quantity). In addition there is a magnetic finite volume shift of 
$\delta m_{\rm mag}$= -0.16 MeV for the proton, giving a total shift of
0.68 MeV. The finite volume corrections for the neutral states are
much smaller- typically less than 0.1 MeV- and have a $1/L^3$ dependence
in the infinite volume limit.
Resonance and Regge region contributions are estimated to be no
larger \cite{GassLeut} than about 30\% of the Born term, so there is 
possibly an error of the order of 0.2 MeV in this result (probably a
conservative estimate, as the higher energy intermediate states correspond
 to shorter distance processes which should be less sensitive to finite
 volume effects). The finite volume
shifts obtained in this way
for each of the isospin splittings in the baryon octet are indicated 
in column 3 of Table 2, and our final estimate (including the
finite volume correction as well as quenched error estimate- see below)
for the  baryon mass in column 5.
 As the L-dependence of these contributions is known,
 calculations on larger lattices will eventually allow a model-independent
 extrapolation to infinite volume.

  Our calculation using quenched gauge configurations neglects graphs with
internal quark loops. Such graphs are known \cite{LangPag,sharpe}
to change the      nonanalytic
$m_q^{3/2}$ dependence in physical baryon masses, by introducing 
additional meson emission and reabsorption processes. 
Such processes also result in
a nonnegligible shift in isospin splittings \cite{leutrev}. For example, 
in the static limit where the nucleon mass is infinite, the effect of the
pion meson clouds surrounding the neutron and proton is to {\em decrease}
the neutron-proton splitting by an amount (in the infinite volume limit)
 0.43$\Delta M_0$, where $\Delta M_0$ is the nucleon splitting in the
 absence of a virtual pion cloud.
 This result is reduced slightly when the meson cloud contribution is
 evaluated fully relativistically- one then obtains 0.41$\Delta M_0$.
 However the shift induced by the meson cloud turns out to be much more
 sensitive to finite volume effects. We shall use the static approximation \cite{leutrev}
 but include the effects of all octet pseudoscalar mesons (assuming SU(3)
 symmetry with a $d:(f+d)$ ratio of 0.62). Discretizing the second order
 shift formula on a LxLxL lattice, one has 
\begin{equation}
\label{eq:mescloud}
 \delta m_{\rm mes}= -C\frac{g_A^2}{f_{\pi}^2}\frac{1}{L^3}
\sum_{\vec{q}}G(q)^2 \frac{\vec{q}^2}{2E(q)} \frac{1}{E(q)+M_2 -M_1}    
\end{equation}
Here, $C$ is an SU(3) Clebsch-Gordon coefficient for emission
and reabsorption  of the meson of mass $m$ from
 a baryon of mass $M_1$, producing a virtual baryon of mass $M_2$.  
$E(q)$ is the virtual meson energy, and $G(q)$
is a dipole form-factor which we take to be $G(q)=\frac{1}{(1+q^2/0.71{\rm GeV}^2)^2}$.
 In general the meson cloud shift includes contributions from quenched 
 nonplanar graphs in the
 cases where the emitted meson only contains valence quarks of the external
 baryon, so these estimates should only be regarded as a rough indication of 
 the magnitude and sign (and probably, an overestimate),
 of the quenched correction. Setting L=12 and
using a lattice scale $a^{-1}$=1370 MeV, together with the quenched masses
from column 2 of Table 1, we obtain the meson cloud shifts given in column 4
of Table 2. The lattice results, corrected for finite volume 
and meson cloud effects, are given in column 5,
and the physical values in column 6. 

\begin{table}
\begin{center}
\caption{Final results for baryon octet splittings 
($\beta$=5.7, 12$^3$x24, 187 configurations).
Errors on lattice results are statistical only.
All results are in MeV.}
\begin{tabular}{|c|c|c|c|c|c|}
\hline
\multicolumn{1}{|c|}{Level}
&\multicolumn{1}{c|}{Raw}
&\multicolumn{1}{c|}{Finite}
&\multicolumn{1}{c|}{Meson}
&\multicolumn{1}{c|}{Total}
&\multicolumn{1}{c|}{Physical}   \\ 
\multicolumn{1}{|c|}{Splitting}
&\multicolumn{1}{c|}{Lattice}
&\multicolumn{1}{c|}{Volume}
&\multicolumn{1}{c|}{Cloud}
&\multicolumn{1}{c|}{Lattice}
&\multicolumn{1}{c|}{Splitting}   \\ \hline
N - P & 2.83 $\pm$ 0.56&  -0.75  &  -0.53  & 1.55 $\pm$ 0.56 &   1.293 \\
$\Sigma^{0}-\Sigma^{+}$  & 3.43 $\pm$ 0.39 & -0.80 &  -0.16 & 2.47 $\pm$ 0.39 &  3.18 $\pm$ 0.1 \\
$\Sigma^{-}-\Sigma^{0}$  & 4.04 $\pm$ 0.36  & +0.86  &  -0.27 & 4.63 $\pm$ 0.36  & 4.88 $\pm$ 0.1\\
$\Sigma^{+}+\Sigma^{-}-2\Sigma^{0}$& 0.61 $\pm$ 0.19& +1.66 &  -0.11 & 2.16 $\pm$ 0.19 & 1.70 $\pm$ 0.15 \\
$\Xi^{-}-\Xi^{0}$  &  4.72 $\pm$ 0.24 & +0.86  &  +0.10  & 5.68 $\pm$ 0.24 & 6.4 $\pm$ 0.6  \\
\hline
\end{tabular}
\end{center}
\end{table}

 The results in Table 2 (which still need to be corrected for finite lattice
spacing effects) suggest
that the baryon isomultiplet splittings prefer a slightly larger up-down quark mass splitting
than obtained from analysis of the pseudoscalar meson spectrum. Indeed the sigma and xi
splittings could be increased by (say) 0.5 MeV without leading to serious disagreement
with the nucleon splitting, given the large statistical error on the latter.
Amusingly, the combination $\Sigma^{+}+\Sigma^{-}-2\Sigma^{0}$, in which linear quark mass
effects cancel, is dominated by finite volume corrections with the present size
lattice. It is evident that the
extremely delicate level of baryon fine structure being considered here
makes essential a detailed study
of all systematic effects, with improved statistics on larger lattices.
Such a study is in progress.

We thank our colleagues in the Fermilab lattice effort and
especially George Hockney for continuing contributions.  
AD gratefully acknowledges the hospitality of the Dept. of Physics and the
support of the EKA Fund,
Columbia University. AD was supported in part by the National Science Foundation
under Grant No. PHY-93-22114.
HBT was supported in part by
the Department of Energy under Grant No.~DE-AS05-89ER 40518.
This work was performed using the ACPMAPS computer
at the Fermi National Accelerator Laboratory, which is operated by
Universities Research Association, Inc., under contract DE-AC02-76CHO3000.



\end{document}